\documentclass[12pt]{article}
\usepackage{epsfig}
\usepackage{color,graphicx}
\usepackage{cite}
\usepackage{amssymb,amsmath}

\newcommand{\be}{\begin{equation}}
\newcommand{\ee}{\end{equation}}
\newcommand{\bea}{\begin{eqnarray}}
\newcommand{\eea}{\end{eqnarray}}

\newcommand{\gsim}{\gtrsim}

\begin{document}
\begin{flushright}
CCTP-2012-02 
\end{flushright}

\begin{center}  

\vskip 2cm 

\centerline{\Large {\bf Magnetic effects in a holographic Fermi-like liquid}}
\vskip 1cm

\renewcommand{\thefootnote}{\fnsymbol{footnote}}

\centerline{
Niko Jokela,${}^{1,2}$\footnote{niko.jokela@usc.es} 
Gilad Lifschytz,${}^3$\footnote{giladl@research.haifa.ac.il} 
and Matthew Lippert${}^4$\footnote{mlippert@physics.uoc.gr}}

\vskip 0.5 cm
\centerline{\it ${}^1$Departamento de  F\'\i sica de Part\'\i culas}
 \centerline{\it Universidade de Santiago de Compostela}
 \centerline{\it and}
 \centerline{\it ${}^{2}$Instituto Galego de F\'\i sica de Altas Enerx\'\i as (IGFAE)}
 \centerline{\it E-15782, Santiago de Compostela, Spain}
 \centerline{\it{ \ }}

\centerline{\it ${}^3$Department of Mathematics and Physics}
\centerline{\it University of Haifa at Oranim}
\centerline{\it Tivon 36006, Israel}
\centerline{\it{ \ }}

\centerline{\it ${}^4$Crete Center for Theoretical Physics}
\centerline{\it Department of Physics}
\centerline{\it University of Crete, 71003 Heraklion, Greece}

\end{center}

\vskip 0.3 cm

\setcounter{footnote}{0}
\renewcommand{\thefootnote}{\arabic{footnote}}

\begin{abstract}
\noindent 
We explore the magnetic properties of the Fermi-like liquid represented by the D3-D7' system.
The system exhibits interesting magnetic properties such as ferromagnetism and an anomalous Hall 
effect, which are due to the Chern-Simons term in the effective gravitational action. 
We investigate the spectrum of quasi-normal modes in the presence of a magnetic field and 
show that the magnetic field mitigates the instability towards a striped phase.
In addition, we find a critical magnetic field above which the zero sound mode becomes massive. 

\end{abstract}

\newpage

\section{Introduction}

In recent years the D3-D7' system has been put forward as a top-down holographic model for fermions in $2+1$ dimensions strongly 
interacting with gauge fields in $3+1$ dimensions \cite{Rey:2008zz,perkraus}. This system has gapped quantum Hall states \cite{bjll} in the 
presence of magnetic fields as well as a rich structure of gapless, metallic states. The stabilized D3-D7' system \cite{bjll} has a large 
number of parameters:  the fluxes $f_1$ and $f_2$ on the two internal spheres, the fermion mass $m$, the charge density $d$, the 
temperature $T$, and the background electric and magnetic fields. At $T=0$, $m=0$, and without any charges or background fields, 
for any choice of fluxes, the system is described by a conformal field theory (CFT) and correlation functions 
have the structure 
consistent with conformal invariance \cite{Davis:2011gi}. As shown in \cite{bnll}, as we turn on charge and 
temperature, the system becomes unstable to form an inhomogeneous phase if 
$\frac{|d|}{T^2}$ is sufficiently large.  
There is, therefore, a quantum critical point (QCP) at the origin of the $(T,d)$-plane.  
When $m \neq 0$ the theory is not conformal and so does not have a quantum critical point; however, 
the structure of its phases and their properties are qualitatively similar. 

In this paper we explore two issues.  One is the effect of an external magnetic field on the gapless phase.  In particular, we find that the magnetic field 
mitigates the instability of the homogeneous phase in the presence of charges and gives a mass to the zero sound mode. 
The second issue is the property of the system above the quantum critical point. We find that the homogenous state exhibits interesting behavior 
at nonzero $m$: an anomalous Hall effect (AHE), resistivity saturation, and ferromagnetism. 
It is interesting to note that the instability, the anomalous Hall effect, and the ferromagnetism all arise from the 
same Chern-Simons term in the effective action.


\section{The D3-D7' Model}

\subsection{Background}

We begin with the near-horizon background of the non-extremal D3-branes:
\begin{eqnarray}
\label{D3metric}
 L^{-2} ds_{10}^2 &=& r^2 \left(-h(r)dt^{2}+dx^2+dy^2+dz^2\right)+
 r^{-2} \left(\frac{dr^2}{h(r)}+r^2 d\Omega_5^2\right) \\
\label{RR_5-form}
F_5 &=& 4L^4\left(r^3 dt\wedge dx\wedge dy\wedge dz\wedge dr 
+  d\Omega_5 \right) \,,
\end{eqnarray}
where $h(r)=1-r_T^4/r^4$ and $L^2=\sqrt{4\pi g_{s} N_3}\, \alpha'$. 
For convenience, we work in dimensionless coordinates, {\em e.g.}, $r=r_{phys}/L$.
This background is dual to ${\cal N}=4$ super Yang-Mills theory at a temperature $T = r_T/(\pi L)$.
We parameterize the five-sphere as an $S^2\times S^2$ fibered over an interval:
\bea
 d\Omega_5^2 &=& d\psi^2 + 
 \cos^2\psi (d\Omega_2^{(1)})^2 + \sin^2\psi (d\Omega_2^{(2)})^2 \nonumber \\
 (d\Omega_2^{(i)})^2 &=& d\theta_i^2 + \sin^2\theta_i d\phi_i \,,
\eea
where $0\leq \psi \leq \pi/2$, $0\leq \theta_i \leq \pi$, and $0\leq \phi_i < 2\pi$.
As $\psi$ varies, the sizes of the two $S^2$'s change. At $\psi=0$ one of the $S^2$'s shrinks to zero
size, and at $\psi=\pi/2$ the other $S^2$ shrinks. The $S^2\times S^2$ at $\psi=\pi/4$ is the ``equator".

\subsection{Probe}

The D7-brane extends in $t,x,y,$ and $r$ and wraps the two two-spheres. 
The D7-brane embedding is then characterized by $\psi(r)$ and $z(r)$. However, excitations 
around this embedding are tachyonic. This instability can be cured by turning on an internal flux \cite{Myers:2008me,bjll}. In our case we turn on fluxes through the two two-spheres 
labeled by the parameters $f_1$ and $f_2$. With the correct choice of $f_1$ and $f_2$, one gets a stable embedding. 
We also consider a nonzero charge density, by including the time component of the worldvolume gauge field $a_0(r)$,
and a background magnetic field $b$.
The D7-brane action has a Dirac-Born-Infeld (DBI) term given by
\bea
\label{DBI_action}
 S_{DBI} & = & -T_7 \int d^8x\, e^{-\Phi} \sqrt{-\mbox{det}(g_{\mu\nu}+ 2\pi\alpha' F_{\mu\nu})} \nonumber \\
 &=& - {\cal N} \int dr\, r^2\sqrt{\left(4\cos^4\psi + f_1^2 \right)
 \left(4\sin^4\psi + f_2^2 \right)}\times \nonumber \\ 
 & & \qquad\qquad \times \sqrt{\left(1+ r^4 h z'^2+ r^2 h \psi'^2 - {a_0'}^2\right)\left(1+\frac{b^2}{r^4}\right)}
\eea
and a Chern-Simons (CS) action given by
\bea
\label{CS_action}
S_{CS} &=& -\frac{(2\pi\alpha')^2T_7}{2} \int P[C_4]\wedge F \wedge F \nonumber\\
&=&  -{\cal N}f_1 f_2 \int dr\, r^4 z'(r)
+ 2{\cal N} \int dr\, c(r) b a_0'(r) \ ,
\eea
where ${\cal N} \equiv 4\pi^2 L^5 T_7 V_{2,1}$
and 
\be
c(r) = \psi(r) - \frac{1}{4}\sin\left(4\psi(r)\right) - \psi_\infty + \frac{1}{4}\sin(4\psi_\infty) \ .
\label{cr}
\ee
Note that $c(r)$, and therefore $\psi(r)$, plays the role of an axion in this model.
One also needs to include boundary terms for the action (see \cite{bjll}), but they do not play any role in this paper.

The asymptotic behaviors of the fields are given by
\bea
\psi(r) & \sim & \psi_{\infty}+mr^{\Delta_{+}}-c_{\psi}r^{\Delta_{-}} \\ 
z(r) &\sim& z_{0} +\frac{f_1 f_2}{r} \\
a_{0}(r) &\sim&  \mu -\frac{d}{r} \ ,
\eea
where the boundary value $\psi_\infty$ and the exponents $\Delta_\pm$ are fixed by the 
fluxes $f_1$ and $f_2$:
\bea
(f_1^2 + 4\cos^4\psi_\infty)\sin^2\psi_\infty =  (f_2^2 + 4\sin^4\psi_\infty)\cos^2\psi_\infty \label{const_solution} \\[5pt]
\Delta_\pm  =  -\frac{3}{2}\pm \frac{1}{2}\sqrt{9+16\frac{f_1^2
+16\cos^6\psi_\infty-12\cos^4\psi_\infty}{f_1^2+4\cos^6\psi_\infty}} \ . \label{deltapm}
\eea
The parameters $m$ and $c_\psi$ correspond to the ``mass" and ``condensate" of the fundamental fermions, respectively, and $\mu$ and $d$ to the 
chemical potential and charge density, respectively.\footnote{The physical charge density is given by
$d_{physical}=8\pi^{3} L^4 \alpha' T_{7} d$.}

The conductivities for the black hole embedding are given by
\bea
\sigma_{xx} & = & \frac{N_3}{2\pi^2}\frac{r_T^2}{b^2 + r_T^4} \sqrt{\tilde{d}(r_T)^2 + (f_1^2+4\cos^4\psi(r_T))(f_2^2+4\sin^4\psi(r_T))(b^2+r_T^4)} \label{sigmaxx_BH} \\
\sigma_{xy} & = & \frac{N_3}{2\pi^2}\frac{j_y}{e} = \frac{N_3}{2\pi^2}\left(\frac{b}{b^2+r_T^4}\tilde{d}(r_T)+2c(r_T)\right) \label{sigmaxy_BH} \ ,
\eea
where $c(r)$ is given by equation (\ref{cr}) and $\tilde d(r) \equiv d -2 bc(r)$. 

Recall that at $b=d=m=T=0$, the system has a quantum critical point. This is reflected by the fact that the induced metric on the D7' brane is that of $AdS_{4}\times S^2 \times S^2 $ (see also \cite{Davis:2011gi}) with the AdS radius given by:
\be
 R_{{\rm{AdS}}}=L\frac{\sqrt{(f_{1}^{2}+4\cos^{4}\psi_{\infty})(f_{2}^{2}+4\sin^{4}\psi_{\infty})}}{\sqrt{(f_{1}^{2}+4\cos^{4}\psi_{\infty})(f_{2}^{2}+4\sin^{4}\psi_{\infty})-f_{1}^{2}f_{2}^{2}}} \ .
\ee

Throughout the paper we will choose for definiteness to present 
the figures with the fluxes $f_1=f_2=\frac{1}{\sqrt 2}$ such that $\Delta_+=-1$, $\Delta_-=-2$, and $\psi_\infty=\frac{\pi}{4}$. 
We will keep the formulas in their general form, however.


\section{Properties of the Fermi-like liquid}

In this section, we further explore the properties of the homogeneous phase away from 
the quantum critical point, at nonzero $T$, $d$, and $m$.  In particular, we study the response to background electric and magnetic fields.

\subsection{Magnetization}\label{sec:magnetization}
The D3-D7' system has some interesting magnetic properties, primarily stemming from the second term in the Chern-Simons action (\ref{CS_action}):
\be
\label{CSterm}
2\mathcal N \int dr c(r) ba'_0(r) \ .
\ee

At the QCP, {\em i.e.}, $b=d=m=T=0$, the system is diamagnetic. 
The combination $m=d=b=0$ gives a semi-trivial\footnote{By semi-trivial we mean a solution to the equations of motion for which $\psi'=0$. 
The function $z'$, however, is non-trivial
as long as both $f_1$ and $f_2$ are nonzero.} embedding: $\psi(r) = \psi_\infty$ and (\ref{CSterm}) vanishes.  
Applying a magnetic field, the dominant contribution to the free energy is the DBI term, and the magnetization $M = -\frac{\partial F}{\partial b}$ is negative.

\begin{figure}[ht]
\center
\includegraphics[width=0.40\textwidth]{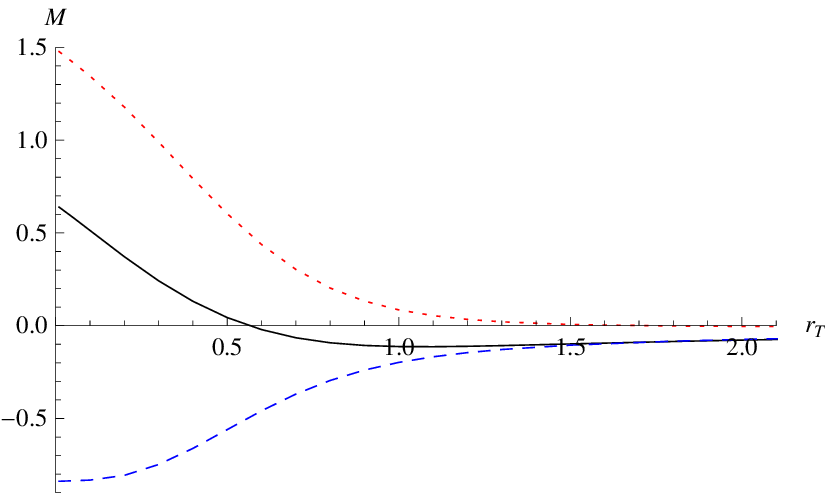}
\includegraphics[width=0.40\textwidth]{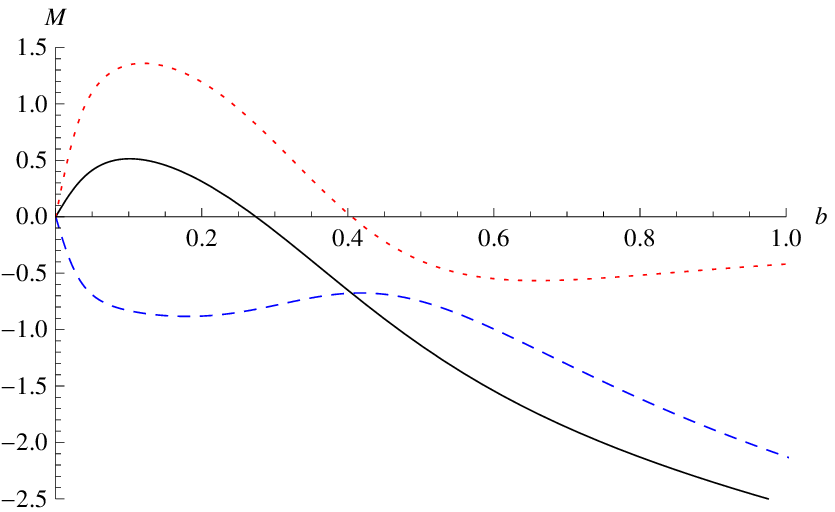}
\caption{Left: The magnetization $M$ versus $r_T$ for $m=0$, $d=1$, and $b=0.1$. Right: The magnetization versus $b$ for $m=0$, $d=1$, and $r_T=0.1$. 
Top curve (dotted red) is the CS contribution, lowest curve the DBI contribution (dashed blue) and the middle curve (solid black) is the total result.}
\label{magnetization}
\end{figure}

Moving away from the QCP, the combination of a nonzero charge and background magnetic field generates a non-trivial embedding, 
even for $m=0$, implying $c(r) \ne 0$ and a non-trivial CS term (\ref{CSterm}).  There are two competing contributions coming from 
the DBI and CS parts of the action. For small $b$, the DBI contributes a negative magnetization while the CS part is positive.  
The relative size of these contributions is temperature-dependent, and as a result, the system is paramagnetic at small 
temperature and diamagnetic at large temperature.  For larger magnetic fields, the CS contribution becomes negative; 
the system is then diamagnetic for all temperatures.  
The magnetization as a function of magnetic field and temperature is shown in Fig.~\ref{magnetization}.

If we further generalize to nonzero $m$, the D3-D7' system becomes ferromagnetic.  
In this case, even for $b=0$, the embedding is necessarily non-trivial, implying that $c(r) \ne 0$.
The Chern-Simons action (\ref{CSterm}) then has a linear term in $b$, which generates a positive magnetization at zero field. 
The DBI action (\ref{DBI_action}) also contributes to the spontaneous magnetization, despite lacking an explicitly linear term; 
the source for the worldvolume gauge field $a_0$ is not $d$ but rather $\tilde{d}=d-2bc(r)$ which, once $a_0$ is integrated out, 
leads terms linear in $b$ and, as it turns out, a negative contribution to $M$ at $b=0$.

\begin{figure}[ht]
\center
 \includegraphics[width=0.40\textwidth]{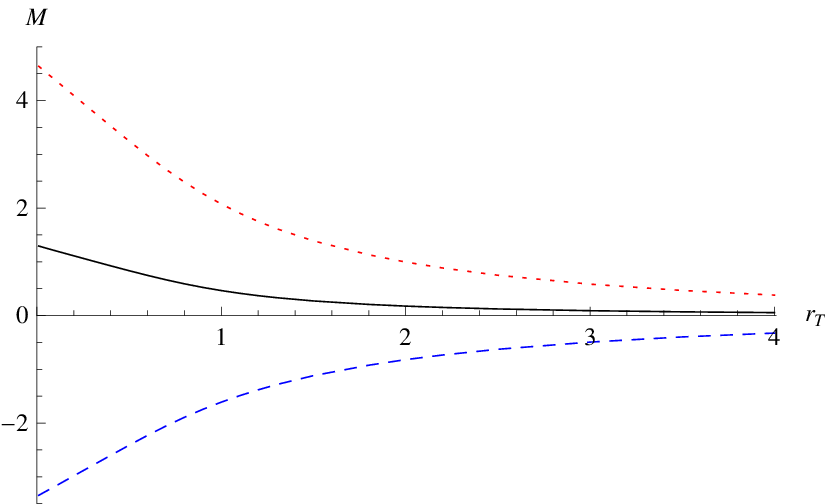}
\includegraphics[width=0.40\textwidth]{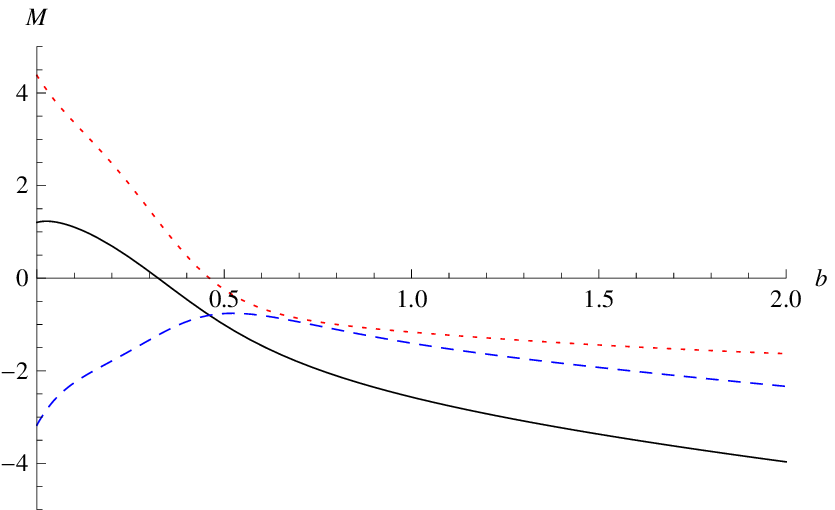}
\caption{Left: The spontaneous magnetization versus $r_T$ for $m=1$, $d=1$, and $b=0$. Right: The magnetization versus $b$ for $m=1$, $d=1$, and $r_T=0.1$. 
Top curve (dotted red) is the CS contribution, lowest curve the DBI contribution (dashed blue) and the middle curve (solid black) is the total result.
}
\label{ferromagnetization}
\end{figure}

Fig.~\ref{ferromagnetization} shows a numerical computation of the magnetization as a function of temperature and magnetic field.  Except for the spontaneous magnetization, {\it i.e.} $M(b=0) \ne 0$, the behavior is qualitatively similar to the $m=0$ case illustrated in Fig.~\ref{magnetization}.

\subsection{Conductivity}
The longitudinal and transverse electrical conductivities were given in general by (\ref{sigmaxx_BH}) and (\ref{sigmaxy_BH}).  
Here we discuss these results in some particular limits.

At the QCP, the embedding is trivial, implying $c(r_T) =0 $ and so $\sigma_{xy} = 0$ as well.  
However, $\sigma_{xx}$ is nonzero, 
which is possible at the QCP, given by
\be
\label{sigmaxxQCP}
\sigma_{xx}=\frac{N_3}{2\pi^2} \sqrt{(f_1^2+4\cos^4\psi_{\infty})(f_2^2+4\sin^4\psi_{\infty})} \ .
\ee

\begin{figure}[ht]
 \center
 \includegraphics[width=0.40\textwidth]{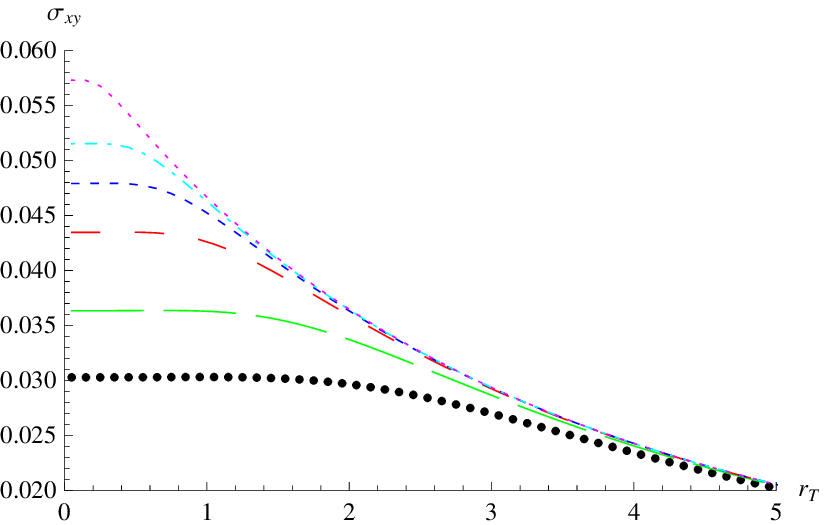}
 \includegraphics[width=0.40\textwidth]{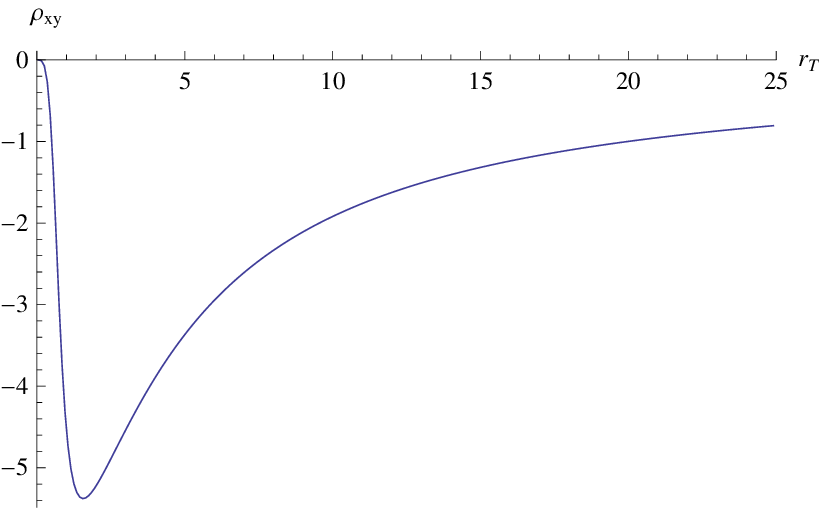}
 \caption{Left: The transverse conductivity $\sigma_{xy}$ at $b=0$ and $m=1$ as a function of temperature $r_{T}$ for various $d=0.1,0.5,1,2,5,10$ (from bottom to top). 
Right: The transverse resistivity $\rho_{xy}$ versus the temperature $r_{T}$ for $b=0$, $d=1$, and  $m=1$.}
\label{nonzerom}
\end{figure}

If we allow $m \neq 0$, we find that the Hall conductivity becomes nonzero, even without a magnetic field.  
The non-trivial embedding implied by $m \neq 0$ leads to a nonzero $c(r_T)$; at $b=0$, the transverse conductivity is then
\be\label{sigmaxyb0}
 \sigma_{xy} = \frac{N_3}{\pi^2}c(r_T) \ .
\ee
Numerical computations of the Hall conductivity and resistivity are shown in Fig.~\ref{nonzerom}.  Such a nonzero 
transverse Hall conductivity at zero magnetic field is called an anomalous Hall effect\footnote{See \cite{ahe} for a review of theoretical and experimental results.} 
and is closely tied to the ferromagnetism noted in Section \ref{sec:magnetization}.

The semiclassical description of the anomalous Hall effect uses transport theory including coherent band mixing. 
This results in a modification for the velocity of the wave packets
\be
\frac{d x^{i}}{dt}=\frac{1}{\hbar}\left[\frac{\partial \epsilon_{n}(k)}{\partial k_{i}}+e{\cal F}^{ij}_{n}E_{j} \right] \ ,
\ee
where $\epsilon_{n}(k)$ is the energy of a Bloch electron in band $n$ and ${\cal F}^{ij}_{n}$ is the Berry curvature in momentum space.
The second contribution on the right hand side gives an extra contribution to the velocity  resulting in a possible nonzero Hall 
conductivity called the intrinsic anomalous Hall conductivity:\footnote{For a nonzero result after summation over bands one 
needs a breaking of time reversal symmetry.}
\be
\sigma_{intrinsic}^{xy}=\frac{e^2}{\hbar (2\pi)^2}\sum_{n}  \int d^2k {\cal F}^{xy}_{n}(k)n_{n}(k, \mu) \ ,
\ee
where $n_{n}(k, \mu)$ are the ground state occupation functions at chemical potential $\mu$ and the integral is over the 
Brillouin zone. If  a band is completely below the Fermi level it contributes an integer to the ``filling fraction'' but a non-quantized 
contribution can come when a band intersects the Fermi surface. In fact, the non-quantized part of the Berry phase contribution 
to the intrinsic Hall conductivity can be written as an integral over the Fermi surface \cite{haldane}. 

We see that this is very similar to what happens in the D3-D7' model.  Here, the CS term contributes an extra 
term to the definition of the current, resulting in the anomalous contribution to the Hall conductivity, 
and $m \neq 0$ breaks the time-reversal symmetry.\footnote{In the case $m=0$ but $b\neq0$, the breaking of time-reversal 
symmetry induces an anomalous term in the Hall conductivity as well, as can be seen in (\ref{sigmaxy_BH}).} Indeed, 
the expression in our brane model for the conductivity at $b=0$, see Eq.~(\ref{sigmaxyb0}), 
is the same expression as for the Hall conductivity in the case of a quantum Hall state. 
The only difference is that here $c(\psi_{T})$ can take a continuous set of values, 
while in the quantum Hall state $\psi_T$ could only be $\pi/2$, so that $c(\psi_T=\pi/2)$ was fixed.

Another notable feature of the D3-D7' model is that in the high-temperature limit the conductivity approaches that of the QCP. 
The high-temperature limit of (\ref{sigmaxx_BH}) and (\ref{sigmaxy_BH}) are
\bea
\sigma_{xx} & \to &  \frac{N_3}{2\pi^2} \sqrt{(f_1^2+4\cos^4\psi(r_T))(f_2^2+4\sin^4\psi(r_T))} \\
\sigma_{xy} & \to &  \frac{N_3}{\pi^2} c(r_T) \ .
\eea
Furthermore, at high temperature $\psi_T \to \psi_\infty$, so the Hall conductivity drops to zero and the longitudinal 
conductivity assumes the QCP form (\ref{sigmaxxQCP}).
\begin{figure}[ht]
\center
\includegraphics[width=0.4\textwidth]{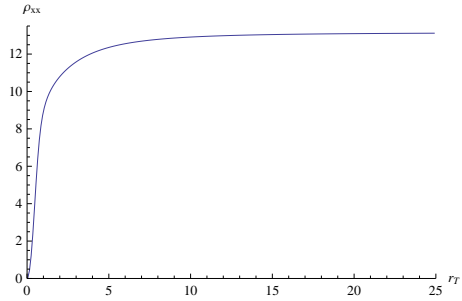}
\caption{The longitudinal resistivity $\rho_{xx}$ as a function of temperature $r_T$ with $d=1$, $m=1$, and $b=0$. 
Although $\rho_{xx}$ initially rises with $r_T$, it eventually saturates at a finite value.}
\label{rhoxxvsrT}
\end{figure}

One consequence of this is that at high temperature the longitudinal resistivity is bounded from above as 
illustrated in Fig.~\ref{rhoxxvsrT}.  This behavior is known as resistivity saturation.\footnote{See \cite{resistivitysaturation} 
for more on the theory and experimental status of the resistivity saturation.} In the usual semiclassical picture, 
resistivity saturation results from a lower bound on the mean free path of the charge carriers. 
In this holographic model, it is instead due to an enhanced pair production rate offsetting the usual temperature suppression of the conductivity.


\section{Quasi-normal mode analysis}

In this section we analyze the fluctuations for the massless $m=0$ background, and, as mentioned before, for the fluxes $f_1=f_2=1/\sqrt 2$. We did not
encounter any qualitative modifications upon taking $m\ne 0$ and thus will not present any pictures for the massive background.

To enter the discussion of the quasi-normal modes, it is convenient to 
switch to a compact radial coordinate. We do so here by inverting it and at the same
time scaling out the dependence of the temperature:
\be
 u \equiv \frac{r_T}{r} \ .
\ee
The addition of a background magnetic field to the setup of \cite{bnll} does not break the rotational symmetry in the $(x,y)$-plane, so we can still restrict to fluctuations 
propagating in the $x$-direction, which schematically take the form $f(u)e^{-i\omega t+i k x}$.

We further define hatted variables and functions such that the temperature is scaled out of all the equations. Most importantly,
\bea
 \hat d & \equiv & \frac{d}{r_T^2} \\
 \hat b & \equiv & \frac{b}{r_T^2} \ .
\eea
We mention in passing that limits $b\to 0$ and $T\to 0$ do not commute, as is common. This noncommutativity manifests itself here in the following way. If one
wishes to consider the zero magnetic field case, $b=0$, then the zero-temperature limit is nothing but $\hat b\to 0$ (and $\hat d\to\infty$), and the 
results in \cite{bnll} follow. However,
if one wishes to keep the physical magnetic field finite and nonzero, the zero temperature limit corresponds to $\hat b\to \infty$.

The original equations of motion for the background fields can be found in \cite{bjll}.  The rescaled equations are \cite{bnll}:
\bea
 \bar z' & = & -\frac{f_1 f_2 h}{\hat g(1+\hat b^2 u^4)} \\
 \bar a'_0 & = & \frac{\hat{\tilde d} h}{\hat g(1+\hat b^2 u^4)} \\
 u^6 \partial_u \left( \hat g (1+\hat b^2 u^4) \psi' \right) &=& -16 u^4 \hat b \bar a_0' \cos^2\psi\sin^2\psi  + \frac{h}{2\hat g} \partial_\psi G \ ,
\eea
where
\bea
 \hat g  & \equiv & \frac{h}{1+\hat b^2 u^4}\sqrt{\frac{\hat{\tilde d}^2 u^4+(1+\hat b^2 u^4)G-h f_1^2 f_2^2}{1+h u^2\psi'^2}} \\
 G & \equiv & \left(f_1^2+4\cos^4\psi\right)\left(f_2^2+4\sin^4\psi\right) \\
 A & \equiv & 1+h u^2\psi'^2+hu^{-4}\bar z'^2-\bar a'^2_0 \ ,
\eea
where the prime denotes differentiation with respect to $u$,   $\bar a_0 '\equiv \partial_r a_0 =  - \frac{u^2}{r_T} \partial_u a_0$, and $\hat{\tilde d}\equiv \hat d-2c(u)\hat b$. Finally 
\be
 \hat m \equiv m r_T^{\Delta_+} = u^{\Delta_+}\left(\psi(u)-\psi_\infty\right)\Big|_{u\to 0} \ ,
\ee
though we will focus here just on massless backgrounds.

\subsection{Fluctuation equations of motion}
In order to obtain the linearized equations of motion for all the fluctuations, one needs to expand the action to second order in fields and their derivatives.
We only consider parametric dependence on the AdS directions. 
The fluctuations are rescaled as follows \cite{bnll}:
\be
\delta\hat z\equiv r_T\delta z, \ \ 
\delta\hat a_{t,x,y}\equiv \frac{\delta a_{t,x,y}}{r_T}, \ \ 
\delta\hat e_x\equiv \frac{\delta e_x}{r^2_T} \equiv \frac{k\delta a_t + \omega\delta a_x}{r_T^2} \ ,
\ee
and the rescaled energy and momentum are defined as $\hat\omega \equiv \omega/ r_T$ and $\hat k \equiv k/r_T$. The equations of motion for the fluctuations can be worked out as in \cite{bnll}, and we will just report the results. In fact, including a magnetic field just adds some extra terms
to the equations, and in the $\hat b\to 0$ limit the equations here collapse to those in \cite{bnll}.

First, for convenience, let us define the function 
\bea
 & & \hat H \equiv \frac{\hat g u^2}{A h}\left(1+\hat b^2 u^4\right)\left(1+hu^{-4}\bar z'^2+hu^2\psi'^2\right)\delta \hat a'_t \nonumber \\
 & & -\left(4\hat b\sin^2(2\psi)+\frac{\hat g}{2h}\bar a'_0\left(1+\hat b^2 u^4\right)\partial_\psi\log G\right)\delta\psi\nonumber\\
 & & +\frac{\hat gu^2}{A}\left(1+\hat b^2 u^4\right)\bar a'_0\psi'\delta\psi' -\frac{\hat g}{Au^2}\left(1+\hat b^2 u^4\right)\bar a'_0 \bar z'\delta\hat z' \ .
\eea

The $\delta\psi$ equation of motion reads:
\bea
 & & \left(-\frac{h}{2\hat gu^4}\left(\partial_\psi^2 G-\frac{1}{2G}(\partial_\psi G)^2\right)+8\hat b\bar a'_0\sin(4\psi)+\frac{u^2}{2}\partial_u\left(\hat g\psi'\left(1+\hat b^2 u^4\right)\partial_\psi\log G\right)   \right)\delta\psi \nonumber\\
 & & = -u^2\partial_u\left(\frac{\hat g}{A}\left(1+\hat b^2 u^4\right)\left(1+hu^{-4}\bar z'^2-\bar a'^2_0\right)\delta\psi'\right)\nonumber\\
 & & +\frac{\hat gu^2}{h^2}\left(-\left(1+\hat b^2 u^4\right)\left(1+hu^{-4}\bar z'^2\right)\hat\omega^2+\left(1+hu^{-4}\bar z'^2-\bar a'^2_0\right)h\hat k^2   \right)\delta\psi\nonumber\\
 & & -\frac{\hat g}{2u^2}\left(1+\hat b^2 u^4\right)\partial_\psi\log G\bar z'\delta\hat z'+\frac{\hat g}{h}\bar z'\psi'\left(-\left(1+\hat b^2 u^4\right)\hat\omega^2+h\hat k^2\right)\delta\hat z\nonumber\\
 & & -u^2\partial_u\left(\frac{\hat g h}{Au^2}\left(1+\hat b^2 u^4\right)\bar z'\psi'\delta\hat z'\right)+\left( 4\hat b\sin^2(2\psi)+\frac{\hat g}{2h}\bar a'_0\left(1+\hat b^2 u^4\right)\partial_\psi\log G   \right)u^2\delta \hat a'_t \nonumber\\
 & & +u^2\partial_u\left(\frac{\hat gu^2}{A}\bar a'_0\psi'\left(1+\hat b^2 u^4\right)\delta\hat a'_t  \right)-\frac{\hat gu^4}{h}\bar a'_0\psi'\hat k\delta \hat e_x\nonumber\\
 & & -i\hat k\left( 4\bar a'_0\sin^2(2\psi)+\frac{h\hat b}{2\hat g\left(1+\hat b^2 u^4\right)}\partial_\psi G-\hat b u^2\partial_u\left(\hat gu^4\psi' \right)  \right)\delta\hat a_y \ .
\label{deltapsiEOM}
\eea

The $\delta z$ equation of motion reads:
\bea
 & & 0 = \frac{\hat g}{h}\bar z'\psi'\left(-\left(1+\hat b^2 u^4\right)\hat\omega^2+h\hat k^2\right)\delta\psi \nonumber\\
 & & -u^2\partial_u\left(\frac{\hat g h}{Au^2}\left(1+\hat b^2 u^4\right)\bar z'\psi'\delta\psi'-\frac{\hat g}{2u^4}\partial_\psi\log G\bar z'\delta\psi\right) \nonumber\\
 &  & -u^2\partial_u\left(\left(1+\hat b^2 u^4\right)\frac{1+hu^2\psi'^2-\bar a'^2_0}{A u^2}\hat g\delta\hat z'\right)\nonumber\\
 & & +\frac{\hat g}{h^2}\left(-\left(1+\hat b^2 u^4\right)\left(1+hu^2\psi'^2\right)\hat\omega^2+\left(1+hu^2\psi'^2-\bar a'^2_0\right)h\hat k^2   \right)\delta\hat z \nonumber\\
 & & +\frac{\hat g}{h}\bar a'_0\bar z'\hat k\delta\hat e_x-u^2\partial_u\left(\frac{\hat g}{Au^2}\left(1+\hat b^2 u^4\right)\bar a'_0\bar z'\delta\hat a'_t\right)\nonumber\\
 & & -i\hat k\hat b u^2\delta\hat a_y\partial_u\left(\hat g\bar z'\right) \ .
 \label{deltazEOM}
\eea

The $\delta a_t$ equation of motion reads:
\bea
 & & 0 = u^2\hat H-\frac{\hat g}{h}u^4\bar a'_0\psi'\hat k^2\delta\psi+\frac{\hat g}{h}\bar a'_0\bar z'\hat k^2\delta\hat z \nonumber\\
 & & -\hat k\frac{\hat g}{h^2}u^4\left(1+hu^{-4}\bar z'^2+hu^2\psi'^2\right)\delta\hat e_x-i\hat k\delta\hat a_y u^2\partial_u\left(2 c(u)-\frac{\hat b\hat g}{h}u^4\bar a'_0\right) \ .
\label{deltaatEOM}
\eea

The $\delta a_x$ equation of motion reads:
\bea
 & & 0 = -\frac{\hat g}{h}u^4\psi'\bar a'_0\hat k\hat\omega\delta\psi+\hat k\hat\omega\frac{\hat g}{h}\bar a'_0\bar z'\delta\hat z-\hat\omega\frac{\hat g}{h^2}u^4\left(1+hu^{-4}\bar z'^2+hu^2\psi'^2\right)\delta\hat e_x\nonumber\\
 & & -i\hat\omega\delta\hat a_y u^2\partial_u\left(2c(u)-\frac{\hat b\hat g}{h}u^4\bar a'_0\right)+\frac{u^2}{\hat\omega}\partial_u\left(\hat g u^2\left(-\delta\hat e'_x+\hat k\delta\hat a'_t\right)\right) \ .
\label{deltaaxEOM}
\eea

The $\delta a_y$ equation of motion reads:
\bea
 & & 0 = i\hat k\hat b\delta\psi u^2\partial_u\left(\hat g u^4\psi'\right)+4i\hat k\bar a'_0\sin^2(2\psi)\delta\psi+i\hat k\frac{h\hat b\partial_\psi G}{2\hat g\left(1+\hat b^2 u^4\right)}\delta\psi \nonumber\\
 &  & +i\hat k\hat b\delta\hat z u^2\partial_u\left(\hat g\bar z'\right)+i\delta\hat e_x u^2\partial_u\left(2c(u)-\frac{\hat b\hat g}{h}u^4\bar a'_0\right)-u^2\partial_u\left(\hat g u^2\delta\hat a'_y\right)\nonumber\\
 & & -\frac{\hat g}{h^2}u^4\left(1+hu^{-4}\bar z'^2+hu^2\psi'^2\right)\hat\omega^2\delta\hat a_y+\frac{\hat g}{h}u^4 A \hat k^2\delta \hat a_y \ .
\label{deltaayEOM}
\eea

And finally, the constraint coming from $\delta a_u$ equation of motion, {\em i.e.}, maintaining the gauge $a_u=0$, reads:
\be
 -\hat\omega\hat H+\frac{\hat k}{\hat\omega}u^2\hat g\left(-\delta \hat e'_x+\hat k\delta\hat a'_t\right) = 0 \ .
\label{deltaauEOM}
\ee

\subsubsection{Decoupling limits}
In general, the equations of motion are completely coupled.  They partially decouple, however, in several different limits. 
For the purposes of this paper, the following cases are relevant:\footnote{Note that our list of decoupling limits is not exhaustive.  
For example, if either of the internal
fluxes $f_1,f_2$ is set to zero, which is the case for Minkowski embeddings, the Chern-Simons term sourcing $\bar z'$ 
vanishes (the first term in (\ref{CS_action})) and $\delta \hat z$ decouples.}

\begin{itemize}
\item {$\hat m=0$ and $\hat d=0$}

In this case $\psi'=0$ and also $\partial_\psi G=0$. The system of equations decouple as follows:
\bea
 \hat k\ne 0 \qquad & : & \qquad (\delta\hat e_x,\delta\psi,\delta\hat a'_t) \perp (\delta\hat z,\delta \hat a_y) \\
 \hat k = 0 \qquad & : & \qquad \delta\hat e_x \perp (\delta\psi,\delta\hat a'_t) \perp \delta\hat z \perp \delta \hat a_y \ .
\eea

\item {$\hat m=0$ and $\hat b=0$}

Again, $\psi'=0$ and $\partial_\psi G=0$. The system of equations now decouple as follows:
\bea
 \hat k\ne 0 \qquad & : & \qquad (\delta\hat z,\delta \hat e_x,\delta\hat a'_t) \perp (\delta\psi,\delta \hat a_y) \\
 \hat k = 0 \qquad & : & \qquad \delta\hat e_x \perp (\delta\hat z,\delta\hat a'_t) \perp \delta\psi \perp \delta \hat a_y \ .
\eea

\item {$\hat k=0$ with $\hat d\ne 0\ne \hat b$}

In this case we no longer have a semi-trivial background ({\em i.e.}, $\psi'\ne 0$), but the scalars decouple from the vectors at vanishing momentum:
\bea
 (\delta\psi,\delta \hat z,\delta\hat a'_t) \perp (\delta \hat e_x,\delta \hat a_y) \ .
\eea
\end{itemize}

\subsubsection{Method of solution}

In this subsection we will briefly recall how the equations of motion are solved to find quasi-normal modes. 
The methodology described here does not essentially differ from that presented in \cite{bnll}, 
so for more details, we refer the reader to \cite{bnll} and
especially to \cite{Amado:2009ts,Kaminski:2009dh}, where the so-called determinant method is explained in depth.
The references \cite{Jokela:2010nu,Jokela:2011sw} consider MN backgrounds, where slight modifications are needed.

Our goal is to find normalizable solutions to the fluctuation equations of motion that have infalling boundary conditions. 
Near the horizon, all the fields have the same singular leading-order 
behavior, $ (1-u)^{\pm i\frac{\hat\omega}{4}}$ (although $\delta\hat a_t$ has an extra factor of $(1-u)$ to guarantee that it vanishes).   
We separate out this leading singular behavior and choose the minus sign in the exponent to obtain the infalling solution. For example,
$\delta\psi= (1-u)^{-i\frac{\hat\omega}{4}}\delta\psi_{reg}$, where 
$\delta\psi_{reg}$ is regular at the horizon.

Furthermore, we demand that the solutions are normalizable near the AdS boundary. 
Here we have five equations of motion (\ref{deltapsiEOM})-(\ref{deltaayEOM}) to solve for five fluctuations 
($\delta\psi_{reg},\delta\hat z_{reg},\delta\hat e_{x,reg},\delta\hat a_{y,reg},\delta\hat a_{t,reg}$), but they are subject to a constraint (\ref{deltaauEOM}). 
Because the equations of motion preserve the gauge condition, as long as they are all satisfied, imposing the constraint is equivalent to imposing it just at the horizon.
Thus, there are two alternative routes one can choose for implementing the constraint: 1) make use of the 
constraint to solve for $\delta\hat a'_t$ in terms of all the other fields, leaving four equations for the four remaining 
fields, or 2) impose the constraint only on the horizon boundary conditions for $\delta\hat a_t'$ and solve all five field equations. Both
of the routes are equivalent, but we found it numerically faster to follow the second path. Therefore, we will not lose any equations but make sure to take the
constraint into account. According to the determinant method, we choose a set of linearly independent boundary conditions at the horizon, namely,
\be\label{eq:bc}
 \{\delta\psi_{reg},\delta\hat z_{reg},\delta\hat e_{x,reg},\delta\hat a_{y,reg}\} = \{(1,1,1,1),(1,1,1,-1),(1,1,-1,1),(1,-1,1,1)\} \ .
\ee
The derivatives at the horizon are set by the equations of motion. They have lengthy expressions, so we do not present them here. 

Finally, for any given momentum we solve the set of differential equations four times, corresponding to the four different boundary conditions in (\ref{eq:bc}). The
interesting object to look at is the determinant
\be
 \det \left.\left( \begin{array}{cccc}
 u^{\Delta_+}\delta\psi_{reg}^I & u^{\Delta_+}\delta\psi_{reg}^{II} & u^{\Delta_+}\delta\psi_{reg}^{III} & u^{\Delta_+}\delta\psi_{reg}^{IV} \\
 \delta\hat z_{reg}^I & \delta\hat z_{reg}^{II} & \delta\hat z_{reg}^{III} & \delta\hat z_{reg}^{IV} \\ 
 \delta\hat e_{x,reg}^I & \delta\hat e_{x,reg}^{II} & \delta\hat e_{x,reg}^{III} & \delta\hat e_{x,reg}^{IV} \\
 \delta\hat a_{y,reg}^I & \delta\hat a_{y,reg}^{II} & \delta\hat a_{y,reg}^{III} & \delta\hat a_{y,reg}^{IV}
\end{array} \right)\right|_{u\to 0} 
\ee
at the AdS boundary. For a given $\hat k$, one then begins to scan over the complex valued energy $\hat\omega$ until a zero of the determinant is found. 
Once this is the case, one concludes that a normalizable solution has been found;
there is a linear combination of the boundary conditions giving the desired normalizable solution,
for which all fluctuations vanish at the AdS boundary.

In practice, we start with a limit of the parameters such that the equations decouple
and consider separately the different fluctuations. The quasi-normal modes 
are identified as the values of $(\hat{\omega},\hat{k})$ where the contribution to the determinant changes sign.
The accuracy of these positions is therefore determined by the resolution of the scan,
which in our case is at least $10^{-3}$.
Away from this limit the determinant is complex in general, and we find the zero of the determinant using Newton's method.

\subsection{Instability at nonzero $b$}

In \cite{bnll} it was shown that in the absence of a magnetic field, the D3-D7' system is unstable if $\hat d = \frac{d}{r_T^2}\gsim 5.5$. 
The true ground state is believed to be a striped phase, resembling a spin and charge density wave. The instabilities
associated with nonzero momenta are quite generic and stem from a Chern-Simons term in the gravitational action. 
Indeed, many other systems with instabilities occuring at some nonzero momentum have been constructed; see \cite{inhomo}.

\begin{figure}[ht]
 \center
 \includegraphics[width=0.40\textwidth]{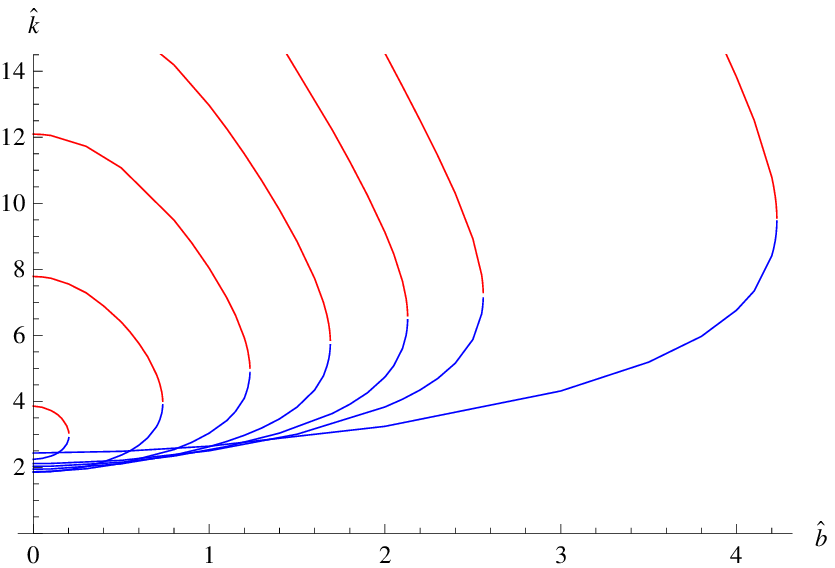}
 \includegraphics[width=0.40\textwidth]{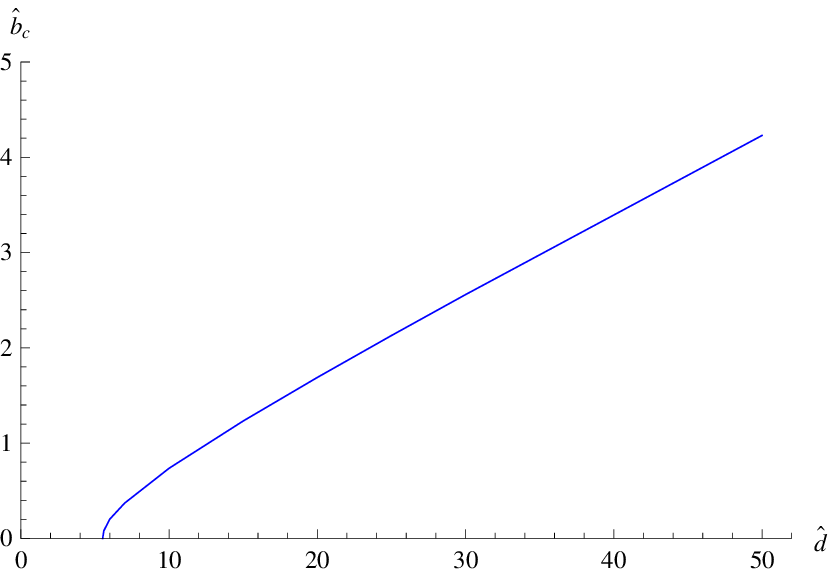}
 \caption{Left: The minimum $\hat k_{min}$ (blue) and maximum $\hat k_{max}$ (red) momentum for the tachyonic instability to occur for the massless embedding $\hat m=0$ 
as a function of the magnetic field for various densities $\hat d=6,10,15,20,25,30,50$ (inside-out). 
Right: The boundary in $(\hat{b}, \hat{d})$ plane, for $\hat m=0$, separating the stable homogeneous state (above) from the unstable 
state (below) where a spin and charge density wave is expected to be the ground state.
Notice the linear behavior of the critical magnetic field for $\hat d\gg 1$.}\label{instability}
\end{figure}

In this section we explore the 
effect of a perpendicular magnetic field, {\em i.e.}, $F_{xy}\propto b$. 
We find that as the magnetic field is increased, stability is enhanced, in the sense that for a fixed charge density the system is stable at a lower temperature. 
The stabilizing effect of the magnetic field can also be seen by looking at the range $(\hat k_{min},\hat k_{max})$ where a tachyon appears; 
as $\hat b$ is increased, this range narrows, as shown in Fig.~\ref{instability}. 
For a given density $d$ at large enough magnetic field $b$, the homogeneous state of the system is stable to an arbitrarily low temperature. 
Fig.~\ref{instability} shows the boundary separating the unstable region (below the curve) from the stable region (above the curve).

\subsection{Fate of zero sound at nonzero $b$}

At zero magnetic field but at nonzero temperature the excitation with the smallest 
damping at low momentum is the purely imaginary hydrodynamical mode. 
As shown in \cite{bnll} this mode meets another purely imaginary longitudinal gauge mode at 
some nonzero momentum, and they become a pair of complex modes which are identified with the positive-temperature zero sound 
modes propagating in opposite directions.\footnote{For a complementary discussion in the supersymmetric D3-D7-brane setup, see \cite{starinets}.}

\begin{figure}[ht]
 \center
 \includegraphics[width=0.30\textwidth]{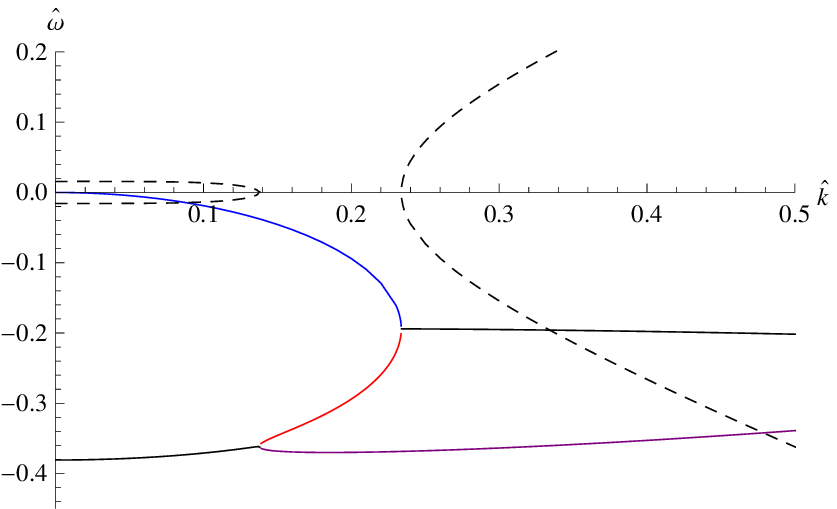}
 \includegraphics[width=0.30\textwidth]{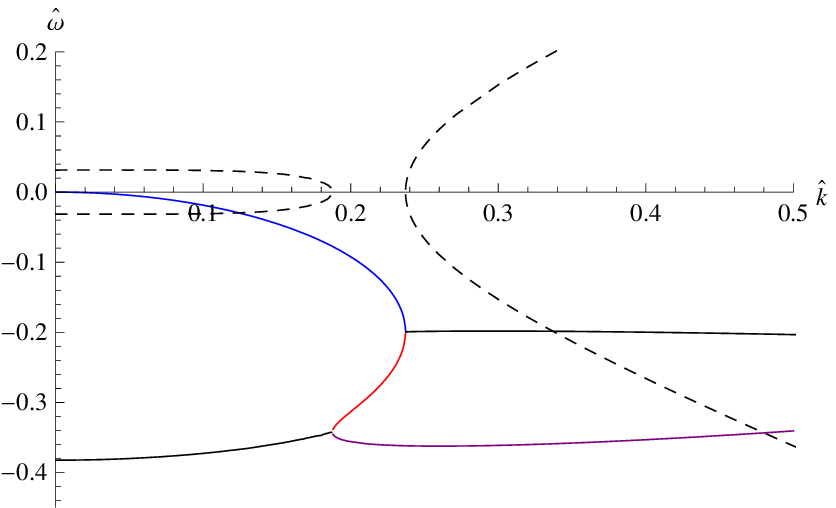}
 \includegraphics[width=0.30\textwidth]{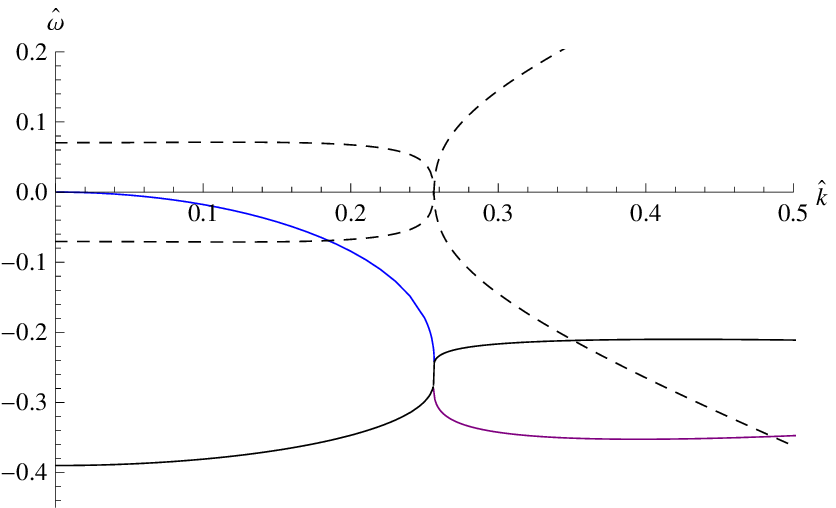}
\caption{The dispersions for the modes closest to the real axis for various $\hat b=0.05$ (left), $0.1$ (middle), 
and $0.22$ (right) for massless background $\hat m=0$ and $\hat d=5$. The purely imaginary hydrodynamical mode is solid blue, and the other purely imaginary modes are solid red and solid magenta.  The complex modes are black, with $\rm{Im} \ \hat\omega$ solid and $\rm{Re} \ \hat\omega$ dashed. The crossover from hydrodynamical to 
collisionless regime, which corresponds to the merging point of the two lowest purely imaginary modes at nonzero momentum $\hat k\sim 0.23\ldots 0.25$, 
is roughly constant as $\hat b$ is varied. 
However, above a critical magnetic field $\hat b_m\approx 0.22$ (right panel), the hydrodynamical mode never merges and the zero sound is massive.}
\label{merging}
\end{figure}

At small enough $\hat b$ this picture persists, except that 
the non-hydrodynamical mode mixes with another purely imaginary transverse gauge mode at small momentum to form two complex modes.  
At larger momentum these modes merge and become two purely imaginary modes, one of which merges at still larger momentum with the hydrodynamical 
mode to become the positive-temperature zero sound modes. 
All this can be seen in Fig.~\ref{merging}. As $\hat b$ is 
increased further, the merging points get closer together and at some critical $\hat b_m$, they coincide. 
From this point on the so-called zero sound mode is lifted and acquires a mass. 

\begin{figure}[!ht]
\center
\includegraphics[width=0.30\textwidth]{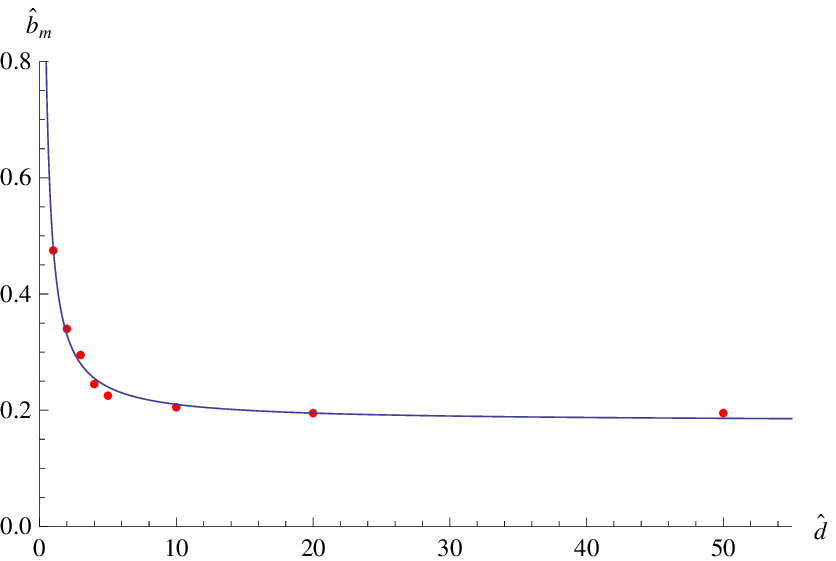}
\includegraphics[width=0.30\textwidth]{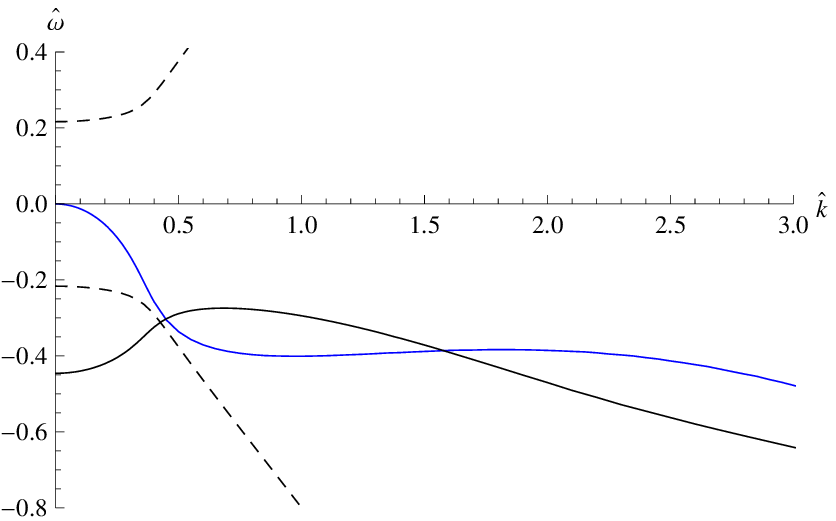}
\includegraphics[width=0.30\textwidth]{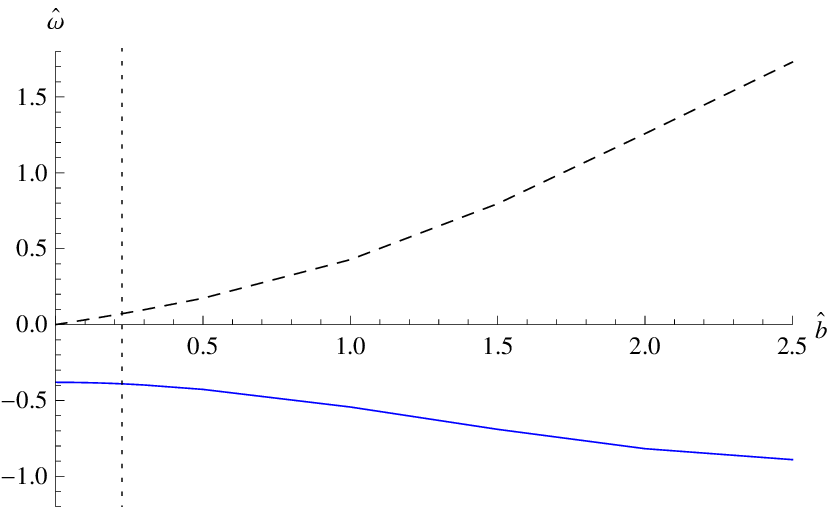}
\caption{Left: The critical magnetic field $\hat b_m$ separating the region with massive zero sound ($\hat b > \hat b_m$) and massless zero sound ($\hat b < \hat b_m$) plotted versus $\hat d$. The dots are 
data for the massless background $\hat m=0$, and the solid curve is the fit $\hat b_m = 0.18+\frac{0.30}{\hat d}$. 
Middle: A typical dispersion of the massive zero sound ($\rm{Im}\  \omega$ is solid black, $\rm{Re} \ \omega$ is dashed black) and the hydrodynamical mode (solid blue) for $\hat m=0$, $\hat d=5$, and $\hat b=0.6>\hat b_c > \hat b_m$. 
Right: The real part (dashed black) and the imaginary part (solid blue) of $\hat\omega(\hat k=0$) for the  mode that will become a part of the massive sound 
mode as a function of $\hat b$ for $\hat d=5$ and $\hat m=0$. The vertical dotted line represents $\hat b_m$.}
\label{zerosound}
\end{figure}

The real part of the now massive sound mode is well approximated by $\hat\omega^2=\hat m_0^2+\hat k^2$.  The critical $\hat b_m$ above which the zero 
sound becomes massive as a function of $\hat{d}$, is shown in Fig.~\ref{zerosound}. 
It is well approximated by $\hat b_m \sim 0.18 +\frac{0.30}{\hat d}$, and a typical dispersion of the massive zero sound for $\hat b>\hat b_c>\hat b_m$ is shown 
in the middle panel of Fig.~\ref{zerosound}. At even larger $\hat k$ (not displayed in Fig.~\ref{zerosound}) the massive zero sound ends up having the smallest 
imaginary part, becoming the dominant mode. The dominance will be slightly more enhanced for larger magnetic fields, where the imaginary parts of the zero sound and
the hydrodynamical mode only cross once.

In Fig.~\ref{zerosound} we also display the real part (the mass of the zero sound) and the imaginary part of the massive zero sound mode at zero momentum.


\section{Conclusion}
In this paper, we have continued our study of the holographic D3-D7' model by investigating the magnetic properties of the ungapped, Fermi-like liquid phase.  
At nonzero mass, the system displays ferromagnetism and an anomalous Hall effect.   
We also found that the longitudinal resistivity saturated at a finite value at high temperature.
We observed that an applied magnetic field has two important effects on the fluctuation spectrum.  
Adding a magnetic field mitigates the modulated instability found at nonzero charge density.  
For given charge density, there is a sufficiently large magnetic field which will render the system stable.
Furthermore, the magnetic field alters the mixing of the quasi-normal modes and, if it is large enough, causes the zero sound mode to acquire a mass.  

One missing element of our investigations is the approach towards the quantum Hall phase.  If one of the internal fluxes  $f_1$ and $f_2$ vanishes, for a specific 
ratio of the magnetic field to the charge density, there is a Minkowski embedding of the D-brane and the fermions become a quantum Hall fluid.  
The quantum Hall fluid is stable and does not suffer from the type of modulated instabilities suffered by the 
ungapped phase \cite{Jokela:2010nu}, but it is unclear how this stabilization comes about.  In this paper, we have worked 
with generic internal fluxes, so the quantum Hall phase was absent.  However, an upcoming work \cite{metal} will extensively address 
these issues in the context of the related D2-D8' model \cite{Jokela:2011eb}.


\bigskip
\noindent

{\bf \large Acknowledgments}

We wish to thank Oren Bergman, Matti J\"arvinen, Daniel Podolsky, Danny Shahar, and Ady Stern for discussions.
N.J. is supported by the MICINN and FEDER (grant FPA2008-01838), the
Spanish Consolider-Ingenio 2010 Programme CPAN (CSD2007-00042), and the Xunta de
Galicia (Conselleria de Educacion and grant INCITE09-206-121-PR). N.J. is
also supported by the Juan de la Cierva program.
N.J. would like to thank the University of Porto and the University of Aveiro for hospitality while this work was in progress.
The research of G.L is supported in part by the Israel Science Foundation under grant no. 392/09, and in part by the National Science Foundation under Grant No. PHY11-25915.
The research of M.L. is supported by the European Union grant FP7-REGPOT-2008-1-CreteHEPCosmo-228644 and 
by the EU program ``Thalis'' ESF/NSRF 2007-2013.  
M.L. would also like to thank the University of Santiago de Compostela for warm hospitality.
N.J. and M.L. also acknowledge the Isaac Newton Institute for Mathematical Sciences, where this work was being completed.

\appendix

\end{document}